\documentclass[aps,prd,superscriptaddress,nofootinbib,11pt]{revtex4}
\usepackage{latexsym,amsmath,amsfonts,amssymb,amstext,mathrsfs,xfrac,esint}
\usepackage[usenames,dvipsnames]{pstricks}

\newcommand{\lbl}[1]{\label{eq:#1}}
\newcommand{ \rf}[1]{(\ref{eq:#1})}

\newcommand{\be}{\begin{equation}}
\newcommand{\ee}{\end{equation}}
\newcommand{\bea}{\begin{eqnarray}}
\newcommand{\eea}{\end{eqnarray}}
\newcommand{\setl}{\setlength\arraycolsep{2pt}}

\newcommand{\noi}{\noindent}
\newcommand{\nn}{\nonumber}
\newcommand{\ra}{\rightarrow}

\newcommand{\cA}{{\cal A}}

\newcommand{\cF}{{\cal F}}
\newcommand{\cG}{{\cal G}}

\newcommand{\cL}{{\cal L}}
\newcommand{\cM}{{\cal M}}

\newcommand{\cO}{{\cal O}}

\newcommand{\cR}{{\cal R}}

\newcommand{\Imm}{\mbox{\rm Im}}
\newcommand{\Ree}{\mbox{\rm Re}}

\newcommand{\MeV}{\mbox{\rm MeV}}
\newcommand{\GeV}{\mbox{\rm GeV}}

\newcommand{\annd}{\mbox{\rm and}}

\begin{document}

\title{Hadronic Vacuum Polarization  Contribution to   $\boldsymbol{g_{\mu}-2}$\\ as a function of  the external lepton mass}

\author{David Greynat}
\email{david.greynat@gmail.com}
\affiliation{No affiliation at present}

\author{Eduardo de Rafael}
\email{EdeR@cpt.univ-mrs.fr}
\affiliation{Aix-Marseille Univ, Universit\'{e} de Toulon, CNRS, CPT, Marseille, France}

\begin{abstract}
The hadronic vacuum polarization contribution to the anomalous magnetic moment of a lepton is considered as a function of the lepton mass. We show how the analyticity properties of this function  allow for its full reconstruction when it is only known in a restricted mass region.  Assuming that LQCD could evaluate it in an optimal mass region we show, within the framework of  a phenomenological  model, how to reconstruct the function  elsewhere, in particular at the muon mass value.
\end{abstract}

\maketitle

\section{Introduction}

{  The latest result for the anomalous magnetic moment of the muon obtained by The Muon g-2 Collaboration at Fermilab~\cite{Fermilab23}:
\be
a_\mu = 116 592 057(25) \times 10^{-11}\quad  (0.21 {\rm ppm})\,,
\ee
when combined with previous measurements at BNL\cite{BNL} and FNAL~\cite{Fermilab21}} brings the present experimental  world average
\be
a_{\mu}({\rm Exp})=116 592 059(22)\times 10^{-11} \,,
\ee
to the remarkable level of a $0.19$ ppm precision. This has to be contrasted with the   theoretical evaluation of $a_{\mu}$ which, at present,  is
not competitive at the same level of precision because of discrepancies in the evaluation of the hadronic vacuum polarization (HVP) contribution.

All data driven determinations of the HVP contribution to the   anomalous magnetic moment of the muon (anomaly for short)  have so far been made  using the   dispersive  representation~\cite{BM61,BdeR68,Gourdin:1969dm}:
\be
\label{eq:BMSR}
a_{\mu}^{\rm HVP}=\frac{\alpha}{\pi}\int_{t_0}^\infty \frac{dt}{t}\underbrace{\int_0^1 dx\  \frac{x^2 (1-x)}{x^2 +\frac{t}{m_{\mu}^2}(1-x)}}_{K(t/{m_{\mu}^2)}}\frac{1}{\pi}\ \Imm\Pi_{\rm had}(t)\,,
\ee
where $\frac{1}{\pi}\Imm\Pi_{\rm had}(t)$ denotes the spectral function of hadronic  vacuum polarization, related by the optical theorem to the  one-photon annihilation cross-section
\be
\sigma(t)_{e^+ e^- \ra {\rm had}} \underset{{m_e\ra 0}}{\thicksim}\frac{4\pi^2 \alpha}{t}\frac{1}{\pi}\Imm\Pi_{\rm had}(t)
\ee
that is obtained from experiments~\cite{Davier},~\cite{Teubner}~\footnote{  See also refs.~[2-5]  in the White Paper~\cite{WP20}. At present, the more recent CMD-3 measurement  of $\sigma(e^+  e^- \ra \pi^+  \pi^-)$ is in tension with previous measurements of this cross-section made by CMD-2~\cite{CMD2}, SND~\cite{SND}, KLOE~\cite{KLOE}, BaBar~\cite{BaBar}, BES~\cite{BES} and CLEO~\cite{CLEO}.}. {  These evaluations of $a_{\mu}^{\rm HVP}$ are to be contrasted with the lattice QCD result reported in ref.~\cite{BMW21}, and subsequently confirmed at least partially in refs.~\cite{L1,L2,L3,L4}. Further discussion on this issue   can be found in ref.~\cite{Davieretal23}.}

Here we shall focus our attention on the  HVP evaluation of  the anomaly when the mass of the external lepton {  $m_l$ has an arbitrary value}, more precisely on the evaluation of the anomaly in terms of the ratio
\be
 \frac{ 4 m_{l}^2}{t_0} \equiv z\quad\ {\rm for}\quad 0\le z \le \infty\,,
\ee
where $m_l$ denotes the external lepton mass  and $t_0$  the fixed hadronic threshold.
 Our aim is to  reconstruct the dependence of the anomaly in a large $z$-domain, covering in particular the muon value, from its  knowledge in a restricted $z$-region. This may be of  interest in the case where one could obtain from {  lattice CQD (LQCD)} simulations accurate determinations  of the anomaly in a given  optimal $z$-region. There is no problem in evaluating the anomaly from {  the same hadronic data} at any desired $z$-value, which offers the possibility of comparing the data $z$-shape to the LQCD  $z$-shape, thus providing a new way of investigating the origin of the HVP discrepancies.

The paper is organized as follows. The HVP anomaly function for an arbitrary lepton mass is introduced in the next section with special attention to its asymptotic behaviour properties. The content of the so called {\it transfer theorem} of Flajolet and Odlyzko~\cite{FO90,FS09} (FO-theorem for short) and its application in implementing  reconstruction approximants to the anomaly function is {  presented} in Section \ref{sec:FO}. An illustration of this reconstruction procedure  is discussed in Section \ref{sec:model} within the framework of a phenomenological model of the hadronic spectral function. This provides a simulation on how  the reconstruction procedure could be applied to LQCD data.

 The conclusions are in Section \ref{sec:conclusions}.

\section{The HVP Anomaly Function}

It is convenient, for our purposes, to use the Mellin-Barnes representation of the  anomaly proposed in ref.~\cite{EdeR14}
\be
\lbl{eq:MBA}
\cA(z)
\equiv  \frac{1}{2\pi i}\int\limits_{c_s-i\infty}^{c_s+i\infty}ds\ z^{1-s}\  \cF(s)\ \cM^{\rm HVP}(s)\,,\quad \cA\left(\frac{4m_{\mu}^2}{t_0} \right)\equiv a_{\mu}^{\rm HVP}\,,\quad\annd\quad  c_s \equiv \Ree (s)\in ]0,1[\,,
\ee
where the function $\cF(s)$ denotes the Mellin transform of the kernel $K(t/m_{l}^2)$ in Eq.~\eqref{eq:BMSR}
\be\lbl{eq:Fs}
\cF(s)=\frac{1}{\sqrt{\pi}}\ \frac{\Gamma(3/2 -s)\ \Gamma(1+s)}{(2-s)\ (3-s)}\ \Gamma(s)\Gamma(1-s)\,,
\ee
and  $\cM^{\rm HVP}(s)$ is the Mellin transform of the {  HVP} spectral function of  hadronic vacuum polarization
\be\label{eq:mellinqcd}
\cM^{\rm HVP}(s)\equiv\int_{t_0}^\infty \frac{dt}{t}\left(\frac{t}{t_0}\right)^{s-1}\frac{1}{\pi}\Imm\Pi_{\rm had}(t)\,,\quad t_0 =4m_{\pi^{\pm}}^2\,, \quad   \Ree(s)\  <1 \,.
\ee
In this representation the $s$-domain of integration in Eq.~\rf{eq:MBA} can then be  extended to the full complex
$s$-plane by analytic continuation.

Moments of the hadronic spectral function correspond to negative integer $s$-values of $\cM^{\rm HVP}(s)$:
\be
\int_{t_0}^\infty \frac{dt}{t}\left(\frac{t_0}{t}\right)^{1+n}\frac{1}{\pi}\Imm\Pi_{\rm had}(t)=\cM^{\rm HVP}(-n)\,,\quad n=0,1,2,\cdots \,.
\ee
They are accessible to experiment and also, in principle, to LQCD because they are related~\cite{deR17} to derivatives of the HVP self-energy function $\Pi(Q^2)$ in the Euclidean at $Q^2 =0$
\be
\label{eq:moments}
\cM^{\rm HVP}(-n)=\frac{(-1)^{n+1}}{(n+1)!} \ (t_0)^{n+1}\left( \frac{\partial^{n+1}}{(\partial Q^2)^{n+1}} \Pi^{\rm HVP}(Q^2)\right)_{Q^2 =0}\,.
\ee
More generally, the Mellin transform  $\cM^{\rm HVP}(s)$  is a meromorphic function with poles in the real $s$-axis at $s=1,2,3,\cdots$. Under the assumption that all the poles are simple, $ \mathcal{M}^{\rm HVP}(s)$ has then a {\it singular expansion}~\cite{FGD94} of the type:
  \be
    \mathcal{M}^{\rm HVP}(s) {\asymp}\  \sum_{k=1}^\infty \frac{(-1)^k r_k} {s-k}\,,
  \ee
	with $(-1)^k r_k$ the residue of the pole at $s=k$.

The function $\cA(z)$ can be used  as a gauge of the underlying HVP structure. {  This is because the kernel $K(t/m_l^2)$ in Eq.\eqref{eq:BMSR}, for different values of the  lepton mass  $m_l$, enhances or depresses the $t$-contribution to  $\cA(z)$ from  {different regions} of  the HVP spectral function. This is illustrated in Fig.~\ref{fig:BMkernel} where the shape of the kernel $K(t/m_l^2)$, normalized to its value at threshold, is plotted as a function of $t$ for  different values of the lepton mass.

\begin{figure}[!ht]
\begin{center}
\includegraphics[width=0.70\textwidth]{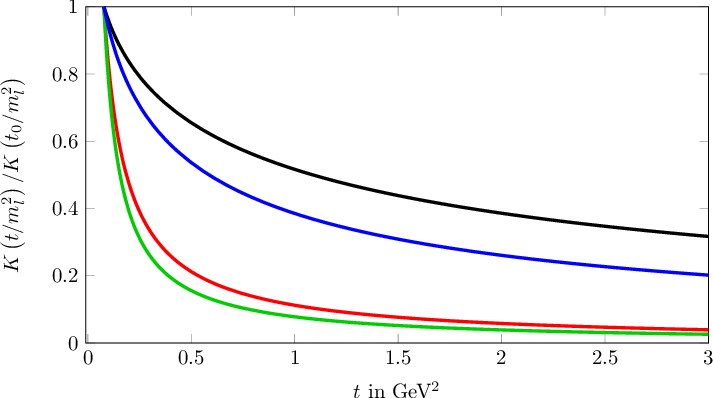} 
\caption{Shape of the kernel $K(t/m_{l}^2)$, normalized to its value at threshold,  for lepton masses: $m_l = m_e$~(green), $m_{\mu}$~(red), $1.0$GeV~(blue) and $1.8$GeV~(black).}\label{fig:BMkernel}
\end{center}
\end{figure}

Two interesting limits of the anomaly function $\cA(z)$ are its behavior at $z\ra 0$ and at $z\ra\infty$.}
 
At $z\ra 0$  the leading contribution is given by the residue of the pole at $s=0$ in the integrand of Eq.~\rf{eq:MBA} with the  result~\cite{BdeR69}:
\be\lbl{eq:exas}
\cA(z)\underset{z\rightarrow 0}{\sim} \frac{z}{12}\ \underbrace{\int_{t_0}^\infty \frac{dt}{t}\ \frac{t_0}{t}\ \frac{1}{\pi}\Imm\Pi^{\rm HVP}(t)}_{\cM^{\rm {HVP}}(0)}= \frac{m_l^2}{3}  \left( -\frac{\partial\Pi^{\rm HVP}(Q^2)}{\partial Q^2}\right)_{Q^2=0} + \cO\left(z^2 \ln z \right)\,.
\ee
This contribution is governed by the long-distance behaviour of HVP expressed here,  either in terms of the  $\cM^{\rm HVP}(0)$ moment of the spectral function  accessible to experiment,   or in terms of  the slope of the HVP self-energy function at the euclidean origin  accessible  to LQCD (see e.g. ref.~\cite{BMW17}). In the case where $m_l \equiv m_{e}$ this asymptotic contribution  practically  amounts to a calculation of the electron anomaly.

On the other hand, at $z\ra\infty$, the leading contribution is given by the residue of the double pole at $s=1$ in the integrand of Eq.~\rf{eq:MBA} with the result
\be\lbl{eq:asfree}
\cA(z)\underset{z\rightarrow \infty}{\sim} \frac{1}{2}\ \frac{  N_c}{3}\left(\sum_{q=u,d,s,c,b,t}e_q^2\right)\ \ln z +\cO({\rm   const.})
\ee
that follows from the familiar short-distance leading behaviour of HVP  in  QCD. In the case where $m_l=m_{\tau}$ it  dominates the HVP contribution to the $\tau$-anomaly.

We want to study the shape of the function $\cA(z)$ in between these two extreme   regimes.  We shall do that   using  the  technique of  reconstruction approximants that follow  from the FO-theorem~\cite{FO90,FS09}, i.e.  the same technique that has been  used in our previous work in refs.~\cite{GdeR22,GdeR23}.
The aim here is to  reconstruct the function $\cA(z)$ in its full $z$-domain when one knows it accurately, e.g. from LQCD simulations, in an optimal $z$-region. Not surprisingly, as earlier discussed in \cite{GP,GMP,GM}, this reconstruction depends crucially on the details of the analytic properties of the function  $\cA(z)$ that we next review.

\subsection{Asymptotic Behaviours of $\cA(z)$ }

The asymptotic behaviour of $\cA(z)$ for $z\le 1$, which includes the cases of the muon and electron $g-2$ in particular,  is governed by the
singular expansion of the $s$-integrand in Eq.~\rf{eq:MBA} at the left of the {\it fundamental strip}~\cite{FGD94}, i.e. the $\Ree(s) \le 0$ region where
\be\lbl{eq:sex}
\cF(s)\cM(s)\underset{\Ree(s)\  \leqslant 0}{\asymp}\  \sum_{p=0}^{\infty}\ \frac{\cL_{p,1}}{(s+p)}+
 \sum_{p=1}^{\infty}\ \frac{\cL_{p,2}}{(s+p)^2}\,,
\ee
and the $\cL_{p,k}$ are the residues of the singularities at $s=-p$  with multiplicities $k=1$ and $k= 2$ generated by the $\cF(s)$ function ($\cM(s)$ is not singular in the region $\Ree(s)\le 0$). The resulting asymptotic expansion of the function $\cA(z)$ has then  the form
\bea\lbl{eq:aslong}
  \mathcal{A}(z) &\underset{z\rightarrow 0}{\sim}  & \sum_{n=0}^\infty \mathcal{L}_{n,1}\  z^{n+1} + \ln z \sum_{n=1}^\infty \mathcal{L}_{n,2}\ z^{n+1} \\
	& = & \mathcal{L}_{0,1}\  z +  \mathcal{L}_{1,2}\  z^2\ln z +  \mathcal{L}_{1,1}\  z^2 + \mathcal{L}_{2,2}\  z^3 \ln z  + \cdots \;,
\eea
where in particular
\be
\cL_{0,1}=\frac{1}{12}\cM^{\rm HVP}(0)\,,\quad \cL_{1,2}=\frac{1}{16}\cM^{\rm HVP}(-1)\,.
\ee
Observe that this series has  non-analytic terms of the type $z^n \ln z$, $n=2,3,4,\cdots$ induced $\underline{only}$ by the double poles of the function $\cF(s)$ in Eq.~\rf{eq:Fs} at $s=-1,-2,-3,\cdots$.

On the other hand the asymptotic expansion of $\cA(z)$ when $z\ge 1$, which applies in particular to the  $g-2$ of the $\tau$-lepton, is governed by the singularities $\underline{both}$ of  $\cF(s)$ $\underline{and}$  $\cM(s)$ at the right of the  fundamental strip in Eq.~\rf{eq:MBA},  i.e. the $\Ree(s) \ge 1$ region where
\be
\cF(s)\cM(s)\underset{\Ree(s)\ge1}{\asymp} \sum_{p=1}^{\infty}\ \  \sum_{k=0,1} \frac{\cR_{p,k}}{(s-p)^{k+1}}+\sum_{p=2,3}\ \frac{\cR_{p,2}}{(s-p)^3}\ +\sum_{p=0}^{\infty}\ \frac{\cR_{3/2 +p,0}}{s-3/2 -p}\,,
\ee
with $\cR_{p,k}$ the residues of the singularities at $s=p$ with  $k+1$ multiplicity generated by the product of the singularities of $\cF(s)$ and $\cM(s)$. The resulting asymptotic expansion of $\cA(z)$ in this case, which also has non-analytic terms,  is slightly more complicated:
\begin{align}\lbl{eq:asshort}
\hspace*{-0.25cm}
  \mathcal{A}(z) &\underset{z\rightarrow \infty}{\sim}-\sum_{n=1}^\infty \mathcal{R}_{n,0}\ z^{-n+1} + \ln z \sum_{n=1}^\infty \mathcal{R}_{n,1}\ z^{-n+1} \nn \\
	& + \frac{1}{2}\ln^2 z  \sum_{n=2,3} \mathcal{R}_{n,2}\ z^{-n+1} - \frac{1}{\sqrt{z}} \sum_{n=0}^\infty \mathcal{R}_{\frac{3}{2}+n,0}\  z^{-n} \\
  & =  \mathcal{R}_{1,1} \ln z - \mathcal{R}_{1,0} -  \mathcal{R}_{\frac{3}{2},0}\frac{1}{\sqrt{z}} + \frac{1}{2}\cR_{2,2}\frac{1}{z}\ln^2 z + \cR_{2,1}\frac{1}{z}\ln z+\cR_{2,0}\frac{1}{z} \cdots \lbl{eq:RSing}\,,
\end{align}
where in particular
\be
\cR_{1,1}\equiv \frac{1}{2}r_{1}=\frac{N_c}{6}\left(\sum_{q=u,d,s,c,b,t}e_q^2\right)\,,\quad\cdots\,.
\ee

 The non-analytic terms that appear in the expansions of $\cA(z)$, both at $z\ra 0$ and at $z\ra \infty$,  are the basic ingredients of the reconstruction approximants of the full $\cA(z)$ function that we next discuss.

\section{The FO-theorem and Reconstruction Approximants}\label{sec:FO}

The FO-theorem~\cite{FO90,FS09} {\it relates the non-analyticity  of a function defined in a finite domain
to the large order behaviour of the coefficients of its Taylor expansion in its analyticity domain.}
In order to apply this theorem to the function $\cA(z)$ we first perform a  mapping of the infinite domain  $0\le z\le \infty$ onto  a finite $\omega$-domain via the conformal  transformation:
\be
z=\left(\frac{1+\omega}{1-\omega}\right)^2 \quad \Longleftrightarrow \quad
\omega= \frac{\sqrt{z}-1}{\sqrt{z}+1}\,,
\ee
which maps the $z$-domain to the unit disc $\vert \omega\vert \le 1$ and in particular:
\be
\begin{array}{lcl}
  z \longrightarrow 0  & \Longleftrightarrow & \omega  \longrightarrow -1 \\
   z \longrightarrow 1 &  \Longleftrightarrow  &\omega \longrightarrow 0 \\
   z \longrightarrow \infty &  \Longleftrightarrow   &\omega \longrightarrow +1 \,.
\end{array}
\ee
The content of the FO-theorem is then encoded in the identity:
\be\lbl{eq:FOid}
\cA\left[z=\left(\frac{1+\omega}{1-\omega}\right)^2 \right] \equiv {\bf A}(\omega) = {\bf A}(0)+
\sum_{n=1}^\infty\  \underbrace{(g_n -g_n^{\rm AS})}_{\cA_n}\ \omega^n \ +\ \underbrace{\sum_{n=1}^\infty\  g_n^{\rm AS}\ \omega^n}_{{\bf A}^{\rm sing}(\omega)}\,,
\ee
where the   $g_n$ denote the coefficients of the Taylor expansion of ${\bf A}(\omega)$ at $\omega\ra 0$ (the image of $z=1$ where $\cA(z)$ is analytic).  The  {  $g_n^{\rm AS}$ are the  coefficients which asymptotically approach the Taylor $g_n$-coefficients} as $n\ra\infty$ and the FO-theorem relates them to the  non-analyticity of the ${\bf A}(\omega)$ function (in our case at $\omega\ra -1$ and at  $\omega\ra +1$, the images of $z\ra 0$ and $z\ra\infty$)   in a precise way. The  third term in the r.h.s. of Eq.~\rf{eq:FOid} is then the  singular function ${\bf A}^{\rm sing}(\omega)$  that emerges from the resulting power series sums.

When $\omega\ra 0$ (the image of $z\ra 1$) the anomaly function ${\bf A}(\omega)$ has  a  Taylor series expansion in $\omega$-powers
\be\lbl{eq:taylor}
{\bf A}(\omega)\underset{\omega\rightarrow 0}{\sim}
 {\bf A}(0) +g_{1}\ \omega + g_{2}\ \omega^2 + g_3 \ \omega^3 + \cdots\,,
\ee
with coefficients:
\be
 {\bf A}(0)  =  \int_{t_0}^\infty \frac{dt}{t}\int_0^1 dx\  \frac{x^2 (1-x)}{x^2 +(1-x)\ \frac{4t}{t_0}}\ \frac{1}{\pi}\ \Imm\Pi_{\rm had}(t)\,, \lbl{eq:gA}
\ee
and
\be
g_n  =   \int_{t_0}^\infty \frac{dt}{t}\int_0^1 dx\ \cG(n,t,t_0)\    \frac{1}{\pi}\Imm\Pi_{\rm had}(t)\,. \lbl{eq:gn}
\ee
\noi
where
\be
\cG(n,t,t_0)=\frac{4 x(1-x)\sqrt{(1-x)\ t\ t_0}}{4(1-x)\ t+x^2\  t_0} \sin\left(n\   {\rm ArcTan}\frac{4 x \sqrt{(1-x)\ t\ t_0}}{   4(1-x)\ t -x^2\ t_0}\right)\,.
\ee
The  coefficients ${\bf A}(0)$ and $g_n$ for $n=1,2,3,\cdots$ are accessible to experiment  by measuring the hadronic integrals above.
In principle they are also accessible to  LQCD from a fit to  evaluations of the anomaly function ${\bf A}(\omega)$ in the region of small $\omega$-values (i.e. around $z=1$), though this   will not be required in the reconstruction procedure that we shall propose. It could  be of interest, however,  if the optimal region for a LQCD evaluation happens to be in the $\omega$-analyticity domain.

We are now in the position to specify the precise way that the FO-theorem relates the coefficients $g_n^{\rm AS}$  to the non-analiticity of the function $\cA(z)$ at $z\ra 0$ ($\omega\ra -1$) and at $z\ra\infty$ ($\omega\ra 1$).

\subsection{Reconstruction Approximants}

Let us  consider the leading non-analytic contributions  in Eqs.~\rf{eq:aslong} and \rf{eq:asshort}.  When expressed in terms of the $\omega$-variable they become
\be
\label{eq:A-1}
  {\bf A}\left(\omega\right)\underset{\omega\rightarrow -1}{\sim}    \frac{1}{8}\mathcal{L}_{1,2}(1+\omega)^{4} \ln(1+\omega)
	 + \cdots \,,
\ee
and
\be
  {\bf A}\left(\omega\right)\underset{\omega\rightarrow 1}{\sim}  -2 \cR_{1,1}\ln(1-\omega) + \cdots \lbl{eq:omplus} \,.
\ee
The FO-theorem relates them  to the  large-$n$ behaviour of the $g_n
^{\rm AS}$ coefficients  as follows:~\footnote{The  mappings for arbitrary types of singular terms in asymptotic expansions are discussed in the Appendices of refs.~\cite{GdeR22},~\cite{GdeR23} and more generally in ref.~\cite{FS09}.}
\be
\frac{1}{8}\  \mathcal{L}_{1,2}\  (1+\omega)^{4}\ln(1+\omega)  \mapsto - \frac{1}{8}\ \cL_{1,2}\ \Gamma(5)\  \frac{(-1)^{n}}{n^5}\,, \lbl{eq:asL}
\ee
and
\be
- 2 \mathcal{R}_{1,1}\ln(1-\omega)  \mapsto    2 \mathcal{R}_{1,1}\frac{1}{n}\,. \lbl{eq:asR}
\ee
The singular functions induced by these  coefficients  are then:
\be
\label{eq:Li5}
-3 \cL_{1,2}\ \sum_{n=1}^\infty  \frac{(-1)^{n}}{n^5}\omega^n = -3 \cL_{1,2}\ {\rm Li}_{5}(-\omega)\,,
\ee
and
\be
-2\cR_{1,1}\   \sum_{n=1}^\infty \frac{1}{n}\omega^n =-2\cR_{1,1}\ln(1-\omega) \,.
\ee
{  The r.h.s. of Eq.~\eqref{eq:Li5}  captures the leading non-analytic contributions of Eq.~\eqref{eq:A-1} but differs from it by regular terms.} Therefore, at the leading order that we are considering,  the singular function  in Eq.~\rf{eq:FOid} becomes:
\be\lbl{eq:singap}
{\bf A}^{\rm sing}(\omega)=  -3\cL_{1,2} \ {\rm Li}_{5}(-\omega)
 -2\cR_{1,1}\ \ln (1-\omega)\,,
\ee
and the underlined $\cA_n$ coefficients in Eq.~\rf{eq:FOid} are
\be
\cA_n =  g_n + 3\cL_{1,2}\frac{(-1)^n}{n^5}+2 \cR_{1,1}\ \frac{1}{n}\,.
\ee

Reconstruction approximants to the anomaly function  ${\bf A}(\omega)$ are then  obtained by restricting  the
 $\omega$-polynomial in Eq.~\rf{eq:FOid} to a finite number of $N$ terms:
\be\lbl{eqfirstap}
{\bf A}(\omega)\approx {\bf A}_{\rm FO}(\omega,N)  ={\bf A}(0) + \sum_{n=1}^N \cA_{n}\ \omega^n +{\bf A}^{\rm sing}(\omega)\,.
\ee
The contribution of the   polynomial in the r.h.s. can in fact  be further improved if expressed in terms of rational approximants~\footnote{See for example ref.~\cite{CV23} and references therein.
Notice  that  rational approximants like $\mathrm{R}^J_K(\omega)$ are different to Pad\'e approximants. They don't  require  convergence order by order in $N$  ($J$ and $K$) and they   don't need the requirement of Stieltjes behaviour of the function $\mathbf{A}(\omega)-\mathbf{A}^\mathrm{sing}(\omega)$. The motivation to introduce them here is based on the content of the FO theorem which implies that in order to fix all the coefficients $\mathcal{A}_n$ one also  needs to know  all the associated singular functions. The rational approximants to the truncated polynomials provides  a way to implement physical requirements that   a limited number of singular functions cannot fully reproduce.}:
{  \be\lbl{eq:rational}
{\bf A}(0)+ \sum_{n=1}^N \cA_{n}\  \omega^n = \frac{\displaystyle \sum_{j=0}^J p_j (1+\omega)^j}{1+\displaystyle\sum_{k=1}^K q_k (1+\omega)^k}\equiv\mathrm{R}^J_K(\omega)\,,\quad\quad J-2+K=N+1\,,
\ee}
with the coefficients $p_j$ and $q_k$  fixed by the following requirements:

\begin{itemize}
	\item The coefficients of the Taylor expansion of $\mathrm{R}^J_K(\omega)$ at $\omega=0$ must coincide with those of the  ${\bf A}(0) + \sum_{n=1}^N \cA_{n}\ \omega^n$ polynomial, which implies
\be
\mathrm{R}^J_K(0)= {\bf A}(0)\quad\annd\quad
\left[\omega^n\right] \mathrm{R}^J_K(\omega) = \cA_{n} \,.
\ee

\item
The $p_j$ and $q_k$ coefficients are furthermore  restricted by the physical requirement that at $\omega=-1$ (i.e. at $t=0$)
\be\lbl{eq:res1}
\mathrm{R}^J_K(-1)= p_0 = - {\bf A}^{\rm sing}(-1)\,,
\ee
and
\be\lbl{eq:res2}
\frac{\partial}{\partial \omega}\mathrm{R}^J_K(\omega)\bigg \vert_{\omega=-1} =\  p_1 - p_0q_1\  =\  - \frac{\partial}{\partial\omega}\mathbf{A}^{\rm sing}(\omega)\bigg \vert_{\omega=-1}\,.
\ee
This  fixes the coefficients $p_0$ and $p_1$ in terms of the other parameters.

\item

With the coefficients $p_j$ and $q_k$ fixed the way described above, the final form of the reconstruction approximants that we shall consider are then
\be\lbl{eq:AJK}
{\bf A}(\omega)\approx {\bf A}_{\rm FO}(\omega,J,K)  \equiv
 \frac{\displaystyle \sum_{j=0}^J p_j (1+\omega)^j}{1+\displaystyle\sum_{k=1}^K q_k (1+\omega)^k} +{\bf A}^{\rm sing}(\omega)\,,
\ee
with ${\bf A}^{\rm sing}(\omega)$ given in Eq.~\rf{eq:singap}.

\end{itemize}

In the next section we shall first check how   these  reconstruction approximants work in the case of a simple phenomenological model of the hadronic spectral function. Encouraged by the results obtained we shall then discuss how to apply them to  LQCD evaluations.

\section{Illustration with a Phenomenological Model}\label{sec:model}

The spectral function of the simple model that we shall consider  is  inspired from $\chi$PT and  phenomenology~\footnote{It is a simplified version of  phenomenological spectral functions discussed in the literature, see e.g. ref.~\cite{CHK21} and references therein.}:
\noi

\be\lbl{eq:SFM}
\frac{1}{\pi}\Imm\Pi^{\rm HVP}_{\rm model}(t)=\frac{\alpha}{\pi}\left(1-\frac{4 m_{\pi}^2}{t}\right)^{3/2} \left\{
\frac{1}{12}\vert F(t)\vert^2 +
\sum_{\rm quarks}e_{q}^2\ \  \Theta(t,t_{c},\Delta)\right\}\theta(t-4m_{\pi}^2)\,.
\ee

\begin{figure}[!ht]
\begin{center}
\includegraphics[width=0.70\textwidth]{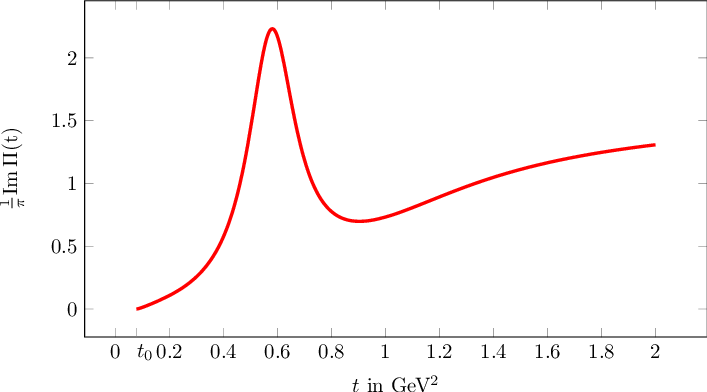}
\caption{The model spectral function in Eq.~\rf{eq:SFM} for $t_c =1~\GeV^2$ and $\Delta=0.5~\GeV^2$ in $\frac{\alpha}{\pi}$-units.}
\lbl{fig:modelsp}
\end{center}
\end{figure}

It has a Breit-Wigner--like modulous squared form factor
\be\lbl{eq:ff2}
\vert F(t)\vert^2=\frac{M_{\rho}^4}{(M_{\rho}^2-t)^2 +M_{\rho}^2\  \Gamma(t)^2}\,,
\ee
with an  energy dependent width:
\be
\Gamma(t)=\frac{M_{\rho} t}{96\pi f_{\pi}^2}\left[\left(1-\frac{4 m_{\pi}^2}{t}\right)^{3/2}\theta(t-4m_{\pi}^2)+\frac{1}{2}\left(1-\frac{4 M_{k}^2}{t}\right)^{3/2}\theta(t-4M_{k}^2)
\right]\,;
\ee
plus a function
\be
\Theta(t,t_{c},\Delta)=\frac{\frac{2}{\pi}\arctan\left(\frac{t-t_{c}}{\Delta}\right)-\frac{2}{\pi}\arctan\left(\frac{t_{0}-t_{c}}{\Delta}\right)}{1-\frac{2}{\pi}\arctan\left(\frac{t_0-t_{c}}{\Delta}\right)}\,,
\ee
with two arbitrary parameters $t_c$ and $\Delta$. This function has been added so as to smoothly match the low energy phenomenological  spectrum of the model  to  the pQCD asymptotic continuum generated by the sum of quark flavors.
The shape of this spectral function, using the  physical central values for $m_{\pi}$, $M_k$,  $M_{\rho}$, $f_{\pi}=93.3~\MeV$, and the choice  $t_c =1~\GeV^2$, $\Delta=0.5~\GeV^2$, with
$\sum_{\rm quarks}e_{q}^2=\frac{5}{3}$,  is shown in Fig~\rf{fig:modelsp}.

{  In this model, the dependence of the integrand of the anomaly in Eq.\eqref{eq:BMSR} normalized to its value a the $\rho-$mass is shown in Fig.~\ref{fig:intanomaly}. One can clearly see how the shape of the integrand is weighted more and more in the energy region above the $\rho$-mass as the external lepton mass $m_l$ increases.
 
\begin{figure}[!ht]
\begin{center}
\hspace*{-1cm}\includegraphics[width=0.70\textwidth]{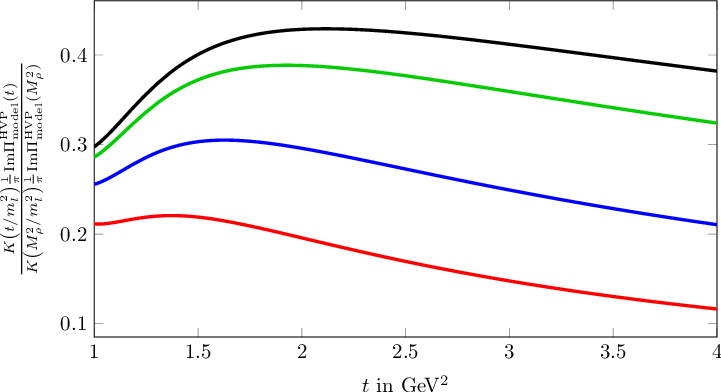} 
\caption{Shape  of the model anomaly integrand in Eq.~\eqref{eq:BMSR},  normalized to its value at the rho mass $(t=M_\rho^2)$, for lepton masses:  $m_l =m_{\mu}$~(red), $M_{\rho}$~(blue), $2 $GeV~(green) and $3 $GeV~(black)}\label{fig:intanomaly}
\end{center}
\end{figure}
}

\subsection{Reconstruction of the Anomaly Function}

The  shape of the anomaly function ${\bf A}(\omega)$ that follows from this phenomenological model, modulated by a convenient factor $(1-\omega)$ that makes it finite everywhere, is shown in Fig.~\rf{fig:Aomega} (the red curve). The Taylor series in Eq.~\rf{eq:taylor} with the model coefficients ${\bf A}(0)$ and $g_n$ evaluated as in Eqs.~\rf{eq:gA} and \rf{eq:gn}    reproduces rather well, with a few terms,  the model function in the region $-0.5\lesssim \omega\lesssim 0.5$. It quickly fails, however,  outside this window. As an example, the dashed green curve shows  the shape of the Taylor series   with seven terms.

The reconstruction approximants improve considerably the Taylor series results. With $N=4$ terms in the $\omega$ polynomial, and the choice $\mathrm{R}_{K=4}^{J=3}$ of the rational improvement in Eq.~\rf{eq:rational}, the resulting reconstruction approximant satisfying the restrictions in Eqs.~\rf{eq:res1} and \rf{eq:res2} is then
\be\lbl{eq:FOAp}
 {\bf A}_{\rm FO}(\omega,3,4) =\frac{\displaystyle \sum_{j=0}^3 p_j (1+\omega)^j}{1+\displaystyle\sum_{k=1}^4 q_k (1+\omega)^k} -3\cL_{1,2} \ {\rm Li}_{5}(-\omega)
 -2\cR_{1,1}\ \ln (1-\omega)\,,
\ee
with

{\setl
\bea
p_0 & = & 2\ \log 2\ \cR_{1,1}\ +\ 3\ \zeta(5)\ \cL_{1,2}\,, \lbl{eq:p0}\\
p_1 & = & \left(-\frac{1}{30}\ \pi^4 \ + \ 3\ \zeta(5)\ q{_1}\right)\ \cL_{1,2} + \left( -1\ +\ 2\ \log2\ q_{1} \right)\ \cR_{1,1}\,. \lbl{eq:p1}
\eea}

\noi
The  values of the parameters of this approximant,  fixed by the model are:
\be
\cL_{1,2}=\frac{1}{16}\cM(-1)=2.86\times 10^{-3}\,,\quad\quad \cR_{1,1}=\frac{5}{6}\,,
\ee
and
\be
p_0=0.0214\,, \quad p_1=-1.6304\,, \quad  p_2=1.6564\,, \quad p_3=-0.3402
\ee
\be
q_1= -1.4698\,, \quad q_2= 0.6070 \quad q_3= -0.0583 \quad q_4= -0.0010\,.
\ee

\begin{figure}[!ht]
\begin{center}
\includegraphics[width=0.6\textwidth]{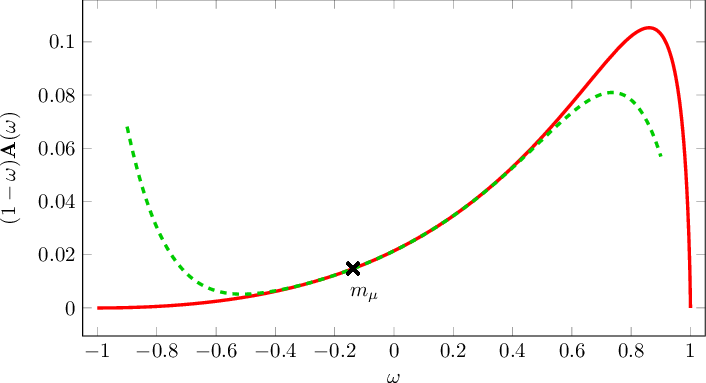}
\caption{Shape  of the function $(1-\omega){\bf A}(\omega)$ of the phenomenological model in red. The black cross in the curve corresponds to the value at the muon mass. The dashed green curve shows the shape of the Taylor series  with $g_{n\le 6}$ terms.}\lbl{fig:Aomega}
\end{center}
\end{figure}

\begin{figure}[!ht]
\begin{center}
\includegraphics[width=0.60\textwidth]{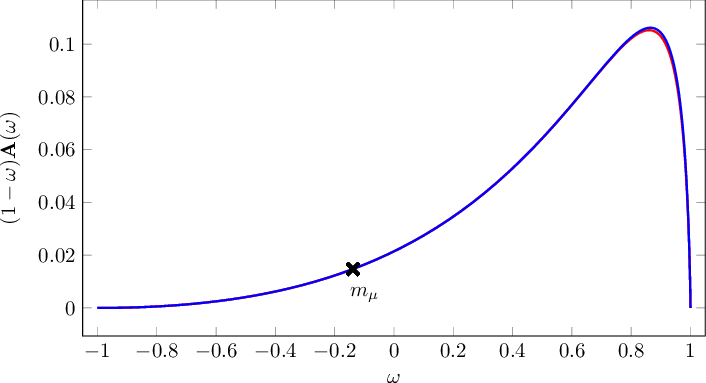}
\caption{ Shape  of the function $(1-\omega){\bf A}(\omega)$ of the phenomenological model in red. The blue curve shows the shape of the reconstruction approximant discussed in the text.}\lbl{fig:4,43}
\end{center}
\end{figure}
\noindent This reconstruction approximant  reproduces very well, as shown in Fig.~\rf{fig:4,43}, the full $\omega$-shape of the model function ${\bf A}(\omega)$:   the blue curve of the reconstruction approximant practically overlaps the red curve of the model function ${\bf A}(\omega)$; in particular,  the value of the anomaly at the physical $m_\mu$-point is reproduced at the level of the number of digits retained in the plots.

\subsection{Reconstruction from a simulation of LQCD  Data}

\begin{figure}[h]
\begin{center}
\includegraphics[width=0.8\textwidth]{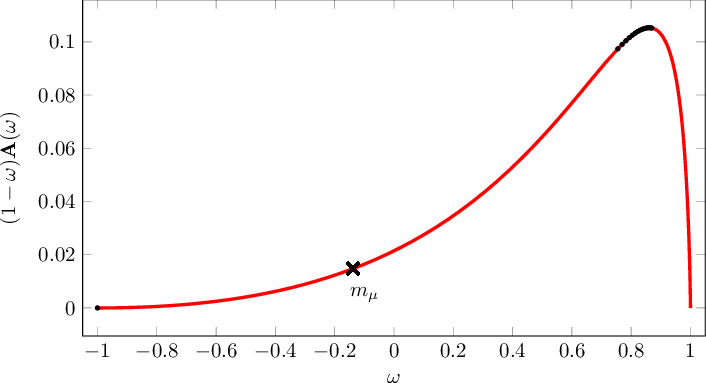}
\caption{The black points are the data input used to reconstruct the full ${\bf A}(\omega)$ function.}\label{fig:points}
\end{center}
\end{figure}

We shall now discuss how to implement  reconstruction approximants in the case where LQCD could provide $z$-data points in an optimal region. To be precise, we assume that the optimal region is at rather large values of the lepton mass~\footnote{This has been suggested to us by Laurent Lellouch.}. We choose the data that simulates potential LQCD data to be  provided by the values of the model  function ${\bf A}(\omega)$ at 15 points equally spaced in the interval $0.76\le\omega\le 0.87$ (i.e. for a lepton mass of $1~\GeV$ to $2~\GeV$); this is shown in Figure (\ref{fig:points}) .
We then fit this data to the coefficients $p_{2,3,4}$ and $q_{1,2,3,4}$ of the function ${\bf A}_{\rm FO}(\omega,4,4)$ in Eq.~\rf{eq:AJK} with the constraints given by Eqs.~\rf{eq:p0} and~\rf{eq:p1} and the coefficients $\cR_{1,1}$ and $\cL_{1,2}$ fixed by the model  (recall that in QCD $\cR_{1,1}$ and $\cL_{1,2}$ are given by the short-distance behaviour of HVP in Eq.~\rf{eq:asfree} and the moment $\mathcal{M}(-1)$ of the HVP self-energy function  in Eq.~\eqref{eq:moments}). The resulting ${\bf A}_{\rm FO}(\omega,4,4)$  approximant is the dashed blue curve in Fig.~\rf{fig:rec34} which reproduces quite well the shape of the  model ${\bf A}(\omega)$ function; in particular the value of the muon anomaly is reproduced at the $0.3\%$ level.

\begin{figure}[h]
\begin{center}
\includegraphics[width=0.8\textwidth]{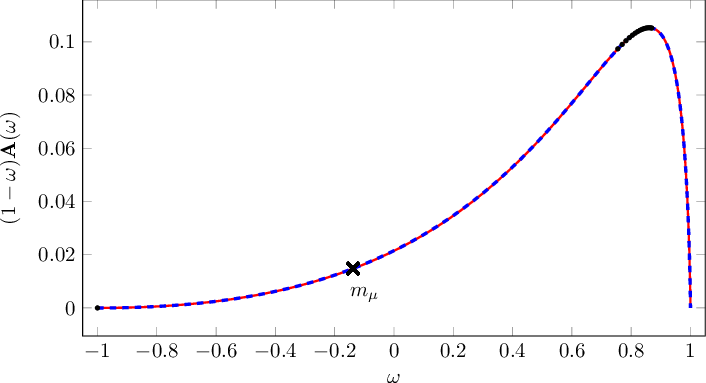}
\caption{Shape of the  reconstruction approximant ${\bf A}_{\rm FO}(\omega,4,4)(\omega)$  in dashed blue. The exact model function is the red curve. The black points are the data used for the fit, the same as in Figure (\ref{fig:points}).}\lbl{fig:rec34}
\end{center}
\end{figure}

\newpage

\section{Conclusions}\label{sec:conclusions}

The function which gives the HVP contribution to the $g-2$ of a lepton in terms of its mass may be a way, via LQCD simulations, to obtain the theoretical value  of the $a_{\mu}^{\rm HVP}$ contribution  to the muon anomaly. We have shown that the  analyticity properties of its functional dependence  on the lepton mass provides a way to reconstruct the full function, in particular at the muon mass value, when one knows its shape in an optimal region of mass values.  Within the example of a phenomenological model of the hadronic spectral function, we have shown how to implement this reconstruction in the case where the  optimal region of lepton masses lies  between $1~\GeV$ and $2~\GeV$. The results are  very encouraging.


\end{document}